\documentclass[conference, 10pt]{IEEEtran} 
\IEEEoverridecommandlockouts

\usepackage{cite}
\usepackage{etoolbox}
\usepackage{array}
\usepackage{soul}
\usepackage[table]{xcolor}
\usepackage{textcomp}
\usepackage{xcolor}
\usepackage{pifont}
\usepackage{setspace}
\usepackage[normalem]{ulem}
\usepackage{hyperref}
\hypersetup{
    colorlinks,
    linkcolor={black},
    citecolor={black},
    urlcolor={blue!80!black}
}
\usepackage{enumitem}
\usepackage[acronym,nohypertypes={acronym,notation}]{glossaries}
\newacronym{3gpp}{3GPP}{3rd Generation Partnership Project}
\newacronym{5g}{5G}{Fifth Generation}
\newacronym{6g}{6G}{Sixth Generation}
\newacronym{aoii}{AoII}{age of incorrect information}
\newacronym{ap}{AP}{access point}
\newacronym{awgn}{AWGN}{additive white Gaussian noise}
\newacronym{ber}{BER}{bit error rate}
\newacronym{bmp}{BMP}{beam-management procedure}
\newacronym{bs}{BS}{base station}
\newacronym{ce}{CE}{channel estimation}
\newacronym{cfo}{CFO}{carrier frequency offset}
\newacronym{cdf}{CDF}{cumulative distribution function}
\newacronym{cfr}{CFR}{channel frequency response}
\newacronym{cir}{CIR}{channel impulse response}
\newacronym{cmW}{cm-Wave}{centimetre waves}
\newacronym{cordic}{CORDIC}{coordinate rotation digital computer}
\newacronym{cp}{CP}{cyclic prefix}
\newacronym{cpu}{CPU}{central processing unit}
\newacronym{crc}{CRC}{cyclic redundancy check}
\newacronym{crdsa}{CRDSA}{contention resolution diversity slotted
ALOHA}
\newacronym{crlb}{CRLB}{Cramér-Rao lower bound}
\newacronym{cto}{CTO}{clock timing offset}
\newacronym{dcc}{DCC}{dynamic cooperation cluster}
\newacronym{dft}{DFT}{discrete Fourier transform}
\newacronym{dl}{DL}{downlink}
\newacronym{dmW}{dm-Wave}{decimetre waves}
\newacronym{dt}{DT}{digital twin}
\newacronym{embb}{eMBB}{enhanced mobile broadband}
\newacronym{ed}{ED}{energy detector}
\newacronym{fdm}{FDM}{frequency-division multiplexing}
\newacronym{fim}{FIM}{Fisher information matrix}
\newacronym{fo}{FO}{frequency offset}
\newacronym{fsa}{FSA}{Framed Slotted-Aloha}
\newacronym{fsm}{FSM}{frequency shifter meta-surface}
\newacronym{fspl}{FSPL}{free-space path loss}
\newacronym{gps}{GPS}{Global Positioning System}
\newacronym{goia}{GOIA}{goal-oriented ISAC access}
\newacronym{goca}{GOCA}{goal-oriented communication access}
\newacronym{htst}{HT-ST}{hollow tone-aided superimposed training}
\newacronym{iap}{IAP}{initial access procedure}
\newacronym{ici}{ICI}{inter-carrier interference}
\newacronym{id}{ID}{identity}
\newacronym{iid}{i.i.d.}{independent and identically distributed}
\newacronym{idft}{IDFT}{inverse discrete Fourier transform}
\newacronym{iot}{IoT}{Internet of things}
\newacronym{isac}{ISAC}{integrated sensing and communication}
\newacronym{isi}{ISI}{inter-symbol interference}
\newacronym{irsa}{IRSA}{Irregular Repetition Slotted Aloha}
\newacronym{jrac}{JR\&C}{joint radar and communication}
\newacronym{lo}{LO}{local oscillator}
\newacronym{los}{LOS}{line-of-sight}
\newacronym{ls}{LS}{least-squares}
\newacronym{mac}{MAC}{medium access control}
\newacronym{mimo}{MIMO}{multiple-input multiple-output}
\newacronym{miso}{MISO}{multiple-input single-output}
\newacronym{mmW}{mmWave}{millimetre waves}
\newacronym{mr}{MR}{maximum ratio}
\newacronym{mse}{MSE}{mean squared error}
\newacronym{nlos}{NLOS}{non-line-of-sight}
\newacronym{nr}{NR}{new radio}
\newacronym{ofdm}{OFDM}{orthogonal frequency division multiplexing}
\newacronym{ofdma}{OFDMA}{orthogonal frequency-division multiple access}
\newacronym{pa}{PA}{power amplifier}
\newacronym{papr}{PAPR}{peak-to-average power ratio}
\newacronym{pbch}{PBCH}{physical broadcast channel}
\newacronym{peb}{PEB}{position error bound}
\newacronym{phy}{PHY}{physical}
\newacronym{pmf}{PMF}{probability mass function}
\newacronym{poia}{POIA}{push-oracle ISAC access}
\newacronym{prach}{PRACH}{physical random access channel}
\newacronym{psam}{PSAM}{pilot symbol assisted modulation}
\newacronym{pss}{PSS}{primary synchronization signal}
\newacronym{qam}{QAM}{quadrature amplitude modulation}
\newacronym{qcqp}{QCQP}{quadratically constrained quadratic program}
\newacronym{qos}{QoS}{quality of service}
\newacronym{ris}{RIS}{reconfigurable intelligent surface}
\newacronym{rb}{RB}{resource block}
\newacronym{re}{RE}{resource element}
\newacronym{se}{SE}{spectral efficiency}
\newacronym{ser}{SER}{symbol error rate}
\newacronym{sic}{SIC}{successive interference cancellation}
\newacronym{sinr}{SINR}{signal-to-interference and noise ratio}
\newacronym{simo}{SIMO}{single-input multiple-output}
\newacronym{snr}{SNR}{signal-to-noise ratio}
\newacronym{soa}{SoA}{state of the art}
\newacronym{speb}{SPEB}{squared PEB}
\newacronym{ss}{SS}{synchronization signal}
\newacronym{ssb}{SSB}{synchronization signal block}
\newacronym{sspa}{SSPA}{solid state power amplifier}
\newacronym{sss}{SSS}{secondary synchronization signal}
\newacronym{st}{ST}{superimposed training}
\newacronym{sto}{STO}{sample timing offset}
\newacronym{tdl}{TDL}{tapped delay line}
\newacronym{tdm}{TDM}{time-division multiplexing}
\newacronym{tdma}{TDMA}{time-division multiple access}
\newacronym{tlst}{TLST}{two layer superimposed training}
\newacronym{to}{TO}{timing offset}
\newacronym{tr}{TR}{tone reservation}
\newacronym{ts}{TS}{training sequence}
\newacronym{uatf}{UatF}{use-and-then-forget}
\newacronym[plural=users equipment]{ue}{UE}{user equipment}
\newacronym{ul}{UL}{uplink}
\newacronym{ula}{ULA}{uniform linear array}
\newacronym{umi}{UMi}{urban micro-cell}
\newacronym{upa}{UPA}{uniform planar array}
\newacronym{urllc}{URLLC}{ultra reliable and low-latency communications}
\newacronym{voi}{VoI}{Value of Information}
\newacronym{viba}{VIBA}{VoI-blind access}
\newacronym{wsn}{WSN}{wireless sensor network}
\newacronym{zf}{ZF}{zero-forcing}
\usepackage{tabularx}
\usepackage{booktabs}
\usepackage{amsfonts,amsmath,amssymb,amsxtra,amsbsy,amsthm}
\usepackage{bbold,bm,cuted,mathtools}
\usepackage{upgreek}
\usepackage[pdftex]{graphicx}
\usepackage{subfig}
\usepackage[font=small]{caption}

\usepackage{tikz}
\usetikzlibrary{patterns,shapes.arrows}
\usepackage{pgfplots}
\usepgfplotslibrary{groupplots,dateplot}
\pgfplotsset{
    compat=newest,
    legend style={font=\tiny, fill opacity=0.7,  draw opacity=1, text opacity=1, draw=white!15!black, legend cell align=left, align=left}, 
    width=6.5cm,     
    yminorticks=false,
    xminorticks=false,
    title style={font=\small},
    label style={font=\scriptsize},
    tick style={color=black},
    tick label style={font=\scriptsize},
    grid style={line width=.1pt, draw=gray!20},
    major grid style={line width=.1pt,draw=gray!20},
}
\DeclareUnicodeCharacter{2212}{−}

\pgfplotstableset{col sep=comma}
\def \fwidth{\columnwidth}
\def \halfwidth{0.53\columnwidth}
\def \fheight {0.42\columnwidth}

\definecolor{lightgray}{HTML}{999999}
\definecolor{color0}{HTML}{00429D}
\definecolor{color1}{HTML}{C3608E}
\definecolor{color2}{HTML}{d27f76}
\definecolor{color3}{HTML}{FFB047}

\newcommand{\T}{^{\mathsf{T}}}     
\renewcommand{\H}{^{\mathsf{H}}}   
\newcommand{\tr}{\mathrm{tr}}
\newcommand{\E}[1]{\mathbb{E}\left[ #1 \right]} 
\newcommand{\mc}[1]{\mathcal{#1}}   
\newcommand{\mb}[1]{\mathbf{#1}}    
\DeclareMathOperator*{\argmax}{arg\,max}    

\newcommand{\ltwonorm}[1]{\left\lVert#1\right\rVert_2} 

\definecolor{gold}{rgb}{0.85,.66,0}
\definecolor{amaranth}{rgb}{0.9, 0.17, 0.31}

\title{Goal-Oriented Access Optimization for ISAC-Enabled Digital Twins}

\author{\IEEEauthorblockN{Fabio Saggese, Federico Chiariotti, Shashi Raj Pandey, Henk Wymeersch, Luca Sanguinetti, Petar Popovski\thanks{F. Saggese (fabio.saggese@ing.unipi.it) and L. Sanguinetti (luca.sanguinetti@unipi.it) are with the Dept. of Information Engineering, University of Pisa, Italy. F. Chiariotti (federico.chiariotti@unipd.it) is with the Dept. of Information Engineering, University of Padova, Italy. H. Wymeersch (henkw@chalmers.se) is with the Dept. of Electrical Engineering, Chalmers University of Technology, Sweden. S. R. Pandey  and P. Popovski (\{srp,petarp\}@es.aau.dk) are with the Dept. of Electronic Systems, Aalborg University, Denmark. F. Saggese's work is funded by Horizon Europe (HE) MSCA Postdoctoral Fellowships, grant~101204088. This work was supported by the Velux Foundation, Denmark, through the Villum Investigator Grant WATER, nr. 37793, and by the HE SNS ``6G-GOALS'', grant 101139232.
}}}

\begin{document}
\maketitle

\begin{abstract}
    \Glspl{dt} of physical systems enable real-time remote tracking, control, and learning, but require to be updated with environmental sensory data 
    to maintain alignment with their physical counterparts. In a network context, \gls{isac} capabilities can expand the \gls{dt}'s environmental awareness by linking received updates to the location where wireless sensors acquired them. 
    Integrating localization services, however, increases the complexity of the communication system, and can only be supported through smart 
    access optimization. To tackle this problem, we design a two-step goal-oriented approach: firstly, sensors with a high \gls{voi} inform the network of their resource demands through a push-based random access; then, pull-based scheduled transmissions of the actual sensory data are optimized to satisfy \gls{isac} performance constraints. This design allows to maximize the \gls{voi} of the information delivered to the \gls{dt} while locating the transmitting nodes, significantly outperforming existing schemes.    
\end{abstract}

\glsresetall

\section{Introduction}

A \gls{dt} of a physical system can enable a wide range of applications, such as experiments, simulations, predictive and counterfactual analyses~\cite{mihai2022digital}, through the replication of its dynamics  in the digital domain; however, it must be maintained up to date with its physical counterpart in real time. This requires strict timeliness from the communication system, as well as a consideration of the most valuable information to maintain its alignment~\cite{liu2025when}. 
This environmental awareness can be supported and expanded by another key technological trend in the evolution of 6G: \gls{isac}, which enables the communication infrastructure to simultaneously act as a sensing platform, measuring properties of the physical environment through their effects on the wireless channel~\cite{kaushik2024toward}. 

There are still several open challenges in the design of \gls{isac} platforms and their integration with \glspl{dt}: providing efficient communication and sensing purposes services requires careful spectrum resource management~\cite{li2023value}.
Efficiently resource allocation for the \gls{dt} to remain up-to-date while guarantee good sensing performance is an even more complex problem~\cite{wang2025optimal}.
Distributed systems such as \glspl{wsn} add another dimension to the problem: while centralized scheduling is always a possible solution, the sensors themselves have access to the information they measure, and may have a better idea of its value to the \gls{dt} application~\cite{topbas2025goal}. 
Unfortunately, at the moment, there is a dearth of effective protocols for \gls{isac} that can maintain \gls{dt} models aligned with a precise localization of \glspl{ue}. 

Most existing works on \gls{isac} networks investigates \gls{jrac} in cooperative scenarios, where a \gls{cpu}, acting as an orchestrator, facilitates the coordination between \glspl{ap}~\cite{wang2025cooperative}. The focus is on letting the \gls{jrac} algorithms feasible, accurate, and scalable~\cite{Liesegang2026unified}, while the relevance of the sensing information fades into the background.

Conversely, the design of goal-oriented \gls{mac} and resource allocation schemes have recently seen significant developments, both adopting a distributed push-based approach~\cite{chiariotti2026theory} in which sensors themselves choose when and whether to transmit their data, and a pull-based approach in which the \gls{cpu} uses its knowledge of the \gls{dt} uncertainty to schedule the sensors that are expected to hold the most useful information~\cite{agheli2026pull}. The combination of push- and pull-based access has been discussed in~\cite{pandey2025medium}, showing that there are significant coexistence gains from integrating the two approaches in \gls{dt} scenarios. However, to the best of our knowledge, goal-oriented access schemes have only considered in communication, without any \gls{isac} capabilities.

This work fills this gap by proposing a joint goal-oriented design of communication and localization resource allocation: we consider a frame structure divided in two subframes~\cite{pandey2025medium}, which allows for sensors with important information to proactively report their \gls{voi} and communication requirements in a \emph{push subframe}, followed by a \emph{pull subframe} in which the \gls{ap} can optimize resource allocation to maximize the total \gls{voi} while respecting localization requirements, in terms of a maximum \gls{peb}. We design a protocol that includes both the grant-free access in the push subframe and the scheduling optimization in the pull subframe and provide analytical bounds for communication and localization performance. Our simulations show that our approach enables the integration of \glspl{dt} and \gls{isac}, significantly outperforming existing \gls{voi}-unaware schemes and approaching oracle-based upper bounds.


\begin{figure}
    \centering
    \includegraphics[width=.8\columnwidth]{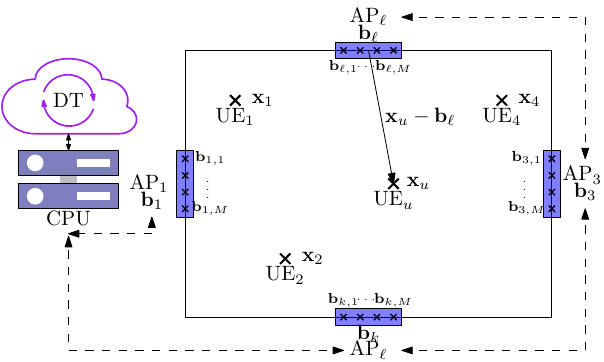}    
    \caption{Example of the scenario.}
    \label{fig:scenario}
\end{figure}

\section{System Model}
\label{sec:model}
We consider the \gls{wsn} scenario of Fig.~\ref{fig:scenario}, with $L$ \glspl{ap}, each equipped with a $M$-antenna \gls{ula}, and $U$ single-antenna mobile \glspl{ue}. The \glspl{ap} are connected through out-of-band channels to a \gls{cpu} running a \gls{dt} of the system.

Assuming a common frame of reference, the position of the antennas of the $\ell$-th \gls{ap} is denoted as $\mb{B}_\ell = [\mb{b}_{\ell,m}]_{m\in\mc{M}} \in\mathbb{R}^{2\times M}$, while its center is $\mb{b}_\ell = \frac{1}{M}\mb{1}_M\T \mb{B}_{\ell}$.
The system works in a \emph{frame-based} fashion, designed so that the position of each \gls{ue}, denoted as $\mb{x}_u\in\mathbb{R}^2$, $\forall u \in\mc{U} = \{1, \dots, U\}$, can be approximated as constant throughout the frame.
The \glspl{ue} have different sensors on board, used to collect sensing data while moving throughout the environment with a maximum velocity $v_u$.
The aim of each node is to send \emph{relevant} sensing data to the \glspl{ap} through a wireless channel. 

\subsection{Application model}

The relevance of observations for the \gls{dt} is assessed through the \gls{voi}. The normalized \gls{voi} of a user $u$ is an \gls{iid} random variable denoted as $V_u\in[0,1]$ characterized by its \gls{cdf} $P_v$. We assume that \gls{ue} $u$ can locally obtain the value of $V_u$~\cite{chiariotti2026theory}, and all \glspl{ue} know $P_v$.

We further assume that the observations obtained by the \glspl{ue} have a considerable size. The observation size for each node, measured in bytes, is an \gls{iid} random variable $B_u$, and it requires significant wireless resources to transmit. Therefore, we consider a two-step system: first, \glspl{ue} with new observations employ a \emph{push} paradigm to inform the \gls{cpu} about their data size and \gls{voi}, then wait to be \emph{pulled} for the actual data transmission, avoiding contention when transmitting data that span multiple packets. The system is designed so that every push transmission attempt indicates that \gls{ue}'s update has a high \gls{voi}, as detailed in Sec.~\ref{sec:mac-push}.

While the \gls{cpu} takes care of processing the receiving signals, it also needs to estimate the position of the \glspl{ue} to link the received data to a specific location in the environment and update the \gls{dt} accordingly. Therefore, the \gls{cpu} will schedule spectrum resources so as to maximize the total \gls{voi}, i.e., optimize communication in a goal-oriented fashion, while guaranteeing an achievable \gls{se} for sensory data delivery and a target \gls{peb} as a proxy metric for localization performance.\footnote{Root mean square error performance may be worse; the \gls{peb} is used to abstract from a specific localization method and its tractability.} Performance guarantees and scheduling are discussed in Secs.~\ref{sec:scheduling-constraints} and~\ref{sec:mac-pull}, respectively.

\begin{figure}
    \centering
    \includegraphics[width=0.75\columnwidth]{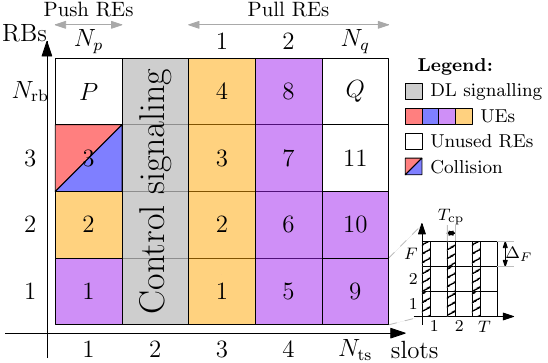}
    \caption{Example of the resource grid for a frame}
    \label{fig:res}
\end{figure}

\subsection{Frame-based OFDM resource grid}
\label{sec:frame}
Each frame, shown in Fig.~\ref{fig:res}, is subdivided into \gls{dl} and \gls{ul} portions: the former is used by the \gls{ap} for control signaling, which is assumed to be ideal throughout the paper, while the latter is reserved for \gls{ue} transmissions.
The frame follows an underlying \gls{ofdm} structure through a time-frequency resource grid of $N_\mathrm{ts}$ time slots and $N_\mathrm{rb}$ \glspl{rb}. A single slot-\gls{rb} is the smallest indivisible unit of resources that can be scheduled, and is referred as a \gls{re}. It comprises $F$ subcarriers spaced by $\Delta_F$ [Hz] and $T$ \gls{ofdm} symbols with \gls{cp} duration of $T_\mathrm{cp}$ [s], assumed longer than channel delay spread. 

The \gls{ul} portion is further subdivided in two subframes.
The \emph{push subframe} occupies the first $N_p$ slots and $N_\mathrm{rb}$ \glspl{rb}, for a total of $P = N_p N_\mathrm{rb}$ \glspl{re}. The \glspl{ue} having relevant data will demand resources by attempting access within the push subframe by transmitting a packet in a \gls{fsa} fashion, as discussed in Sec.~\ref{sec:mac-push}.
This packet only contains the current \gls{voi}, $V_u$, and the observation size, $B_u$, and thus it may be transmitted in a single \gls{re}. 
The \emph{pull subframe} occupies $N_q$ time slots and $N_\mathrm{rb}$ \glspl{rb}, for a total of $Q = N_q N_\mathrm{rb}$ \glspl{re}, with $Q \ge P$. After decoding the packets in the push subframe, the \gls{ap} is able to find a pull-based scheduling policy satisfying the communication and localization criteria through the procedure presented in Secs.~\ref{sec:scheduling-constraints} and~\ref{sec:mac-pull}.
The scheduled \glspl{ue} employ the \glspl{re} for the transmission of the relevant data, while the \gls{ap} exploits the channel estimates in those \glspl{re} to perform positioning on the transmitting \gls{ue}. We denote as $\mc{P}$ and $\mc{Q}$ the set containing the indexes of the \glspl{re} of the push and pull subframes, respectively. 

The \gls{dl} \emph{control signaling} portion occupies the remaining time slots between the two subframes, and it is used for \emph{broadcasting} the schedule of the pull subframe.

\subsection{PHY layer modeling}
\label{sec:signal}
We assume perfect synchronization between the \gls{ap}'s \glspl{lo}~\cite{demir2021cellfree, Liesegang2026unified}. The \glspl{ue} perform synchronization employing the 5G \gls{nr} standard procedure on the \gls{prach}~\cite{3gpp:access:protocol}, compensating their \gls{lo} up to a residual. The residual \glspl{fo} caused by \gls{cfo} and the Doppler effect is considered negligible due to 1) higher precision for \gls{cfo} compensation~\cite{Omri2019:synch}, 2) low speed $v_u$, and 3) positioning focus.
Conversely, we denote the residual \gls{cto} of \gls{ue} $u\in\mc{U}$ at \glspl{re} as $\delta\tau_u$, assumed to be approximately constant throughout the frame.  

For the channel modeling, let us consider a single \gls{re}, $i$, and a single \gls{ue}, $u\in\mc{U}$. The sets indexing the subcarriers and the \gls{ofdm} symbols in \gls{re} $i$ are $\mc{F}_i$ and $\mc{T}_i$. 
We assume a \gls{los} dominant channel model. Thus, the (concatenation of the) \gls{ul} channel vectors between \gls{ue} $u$ and the $L$ \gls{ap} varies only over the subcarriers and is given by~\cite{Wymeersch2022sensingsota, Wu2024sensing-asynch}:
\begin{equation}\label{eq:channel-model-full}
\mb{h}_{u,i}  = \sum_{\ell\in\mc{L}} \sqrt{\beta_{u,\ell}} e^{j\psi_{u,\ell}} \,  (\mb{e}_\ell \otimes \mb{f}_{u,\ell,i} \otimes \mb{a}_{u,\ell}) \in\mathbb{C}^{LFM}.
\end{equation}
where $\mb{e}_\ell = [\mb{0}_{\ell-1}\T, 1, \mb{0}_{L-\ell}\T]\T \in\mathbb{R}^{L}$; $\bm{\beta}_u = [\beta_{u,\ell}]_{\ell\in\mc{L}}$ is the large scale fading, and $\bm{\psi}_u =[\psi_{u,\ell}]_{\ell\in\mc{L}}$ is the phase offset~\cite{Wymeersch2022sensingsota}; vector
$\mb{f}_{u,\ell, i} \in\mathbb{C}^{F} = [\exp({-j2\pi f \Delta_F \tau_{u,\ell}})]_{f\in\mc{F}_i}\T$ accounts for the effect of delay due to propagation and \gls{cto}, $\tau_{u,\ell} = \ltwonorm{\mb{x}_u - \mb{b}_\ell} / c + \delta\tau_{u,\ell}$, with $c$ being the speed of light; $\mb{a}_{u,\ell}\in\mathbb{C}^{M} = [\exp( j\frac{2\pi}{\lambda} (\mb{b}_{\ell, m} -  \mb{b}_\ell)\T(\mb{x}_{u} - \mb{b}_\ell) / \ltwonorm{\mb{x}_u - \mb{b}_\ell})]_{m\in\mc{M}}$ is the steering vector with $\lambda$ being the wavelength. 
%

The signal received for a singleton \gls{re}, i.e., an \gls{re} over which only a single \gls{ue} transmits, is
\begin{equation} \label{eq:received-signal}
    \mb{Y}_i \in\mathbb{C}^{LFM \times T} = \sqrt{p_u} \mb{h}_{u,i} (\mb{1}_{LM} \otimes \mb{S}_{u,i}) + \mb{W}_i,
\end{equation}
where $\mb{S}_{u,i}\in\mathbb{C}^{F \times T}$ is the block of  transmitted symbols  with $\E{\mb{S}_{u,i}}=\mb{0}_{F \times T}$ and $|(\mb{S}_{u,i})_{t,f}|^2 = 1$; $p_u$ is the transmission power of \gls{ue} $u\in\mc{U}$; $(\mb{W}_i)_{t\in\mc{T}} \sim \mc{CN}(\mb{0}_{LM}, \sigma_w^2 \mb{I}_{LMF})$ is the \gls{awgn} of the \glspl{ap}.

To perform \gls{ce} at the \gls{ap} side, each first \gls{ofdm} symbol in the \gls{re}, $(\mb{S}_{u,i})_{:,1}$, carries a known pilot symbol. Employing \gls{ls} estimation yields~\cite{demir2021cellfree}
\begin{equation} \label{eq:ce}
    \hat{\mb{h}}_{u,i} = p_u^{-\frac{1}{2}} (\mb{Y}_i)_{:,1} \odot (\mb{S}_{u,i}^*)_{:,1}  = \mb{h}_{u,i} + \tilde{\mb{w}}_i,
\end{equation}
where $\tilde{\mb{w}}_i = (\mb{W}_i)_{:,1} / \sqrt{p_u} \sim \mc{CN}(\mb{0}_{LFM}, \frac{\sigma_w^2}{p_u}\mb{I}_{LFM})$. If the \gls{re} $i$ is not a singleton, e.g.,  due to  a collision within the push subframe $i\in\mc{P}$, the \emph{\gls{ce} is unfeasible due to pilot contamination} and the data transmitted in the \gls{re} is lost.\footnote{This is the worst-case scenario: due to the capture effect, data from \glspl{ue} with favorable propagation conditions may still be decoded.} Within the pull subframe, scheduling is made in an orthogonal manner preventing pilot contamination--see Sec.~\ref{sec:scheduling-constraints}. 

\section{Goal-Oriented Access Design}
\label{sec:design}

This section outlines the design strategies to maximize the total \gls{voi} per frame. The proposed approach ensure a per-\gls{ue} reliable delivery of the sensory data of size $B_u$, and a \gls{peb} lower than $\epsilon$ by allocating the $Q$ \glspl{re} within the pull subframe.

\subsection{Push subframe strategy}
\label{sec:mac-push}

The \glspl{ue} apply a threshold policy during the push subframe, transmitting a message over a randomly selected \gls{re} among $\mc{P}$ if $V_u>\theta$, which maximizes the expected \gls{voi} under our assumptions~\cite{chiariotti2026theory}. As all \glspl{ue} have the same threshold, the transmission probability of each \gls{ue} is $1-P_v(\theta)$, and the probability that $n$ \glspl{ue} will attempt to reserve resources is
\begin{equation}
    p_{\text{tx}}(n|\theta)=U!(n!(U-n)!)^{-1}P_v(\theta)^{U-n}(1-P_v(\theta))^n.
\end{equation}
Following the signal model in the previous section, we consider the success probability for two or more \glspl{ue} colliding over the same \gls{re} to be negligible, and the erasure probability for a singleton \gls{re}, to also be negligible.

We define $\mc{C}(k, n)=\left\{\mb{c}\in\mathbb{Z}^{k}:c_i\geq2\,\forall i\wedge\sum_{i=1}^{k}c_i=n\right\}$ as the set of possible combinations of $n$ \glspl{ue} colliding over $k$ \glspl{re}. The probability of having $s$ singleton \glspl{re} and $c$ collided \glspl{re}, given that $n$ nodes transmit, is then
\begin{equation}
    p_o(s,c|n)=\begin{cases}
\frac{\binom{n}{s} \sum_{\mb{k}\in\mc{C}(c,n-s)}\binom{n-s}{\mb{k}}P!}{P^n(P-s-c)!c!}, &c>0, n\geq s+2c;\\
    \frac{P!}{P^s(P-s)!}&c=0,s=n;\\    
    0, &\text{otherwise.}
    \end{cases}
\end{equation}
where $\binom{n-s}{\mb{k}}$ is the multinomial coefficient. Naturally, the maximum value of $s$ is $P$.
The probability of $s$ successful transmissions can then simply computed using the law of total probability as
$p(s|\theta)=\sum_{n=s}^U  \sum_{c=0}^{\min\left(\frac{n-s}{2},P-s\right)}p_o(s,c|n)p_{\text{tx}}(n|\theta)$. It is thus possible to optimize the push subframe. Regardless, we denote the \emph{set of \glspl{ue} that transmits successfully} as $\mc{S}\subseteq{\mc{U}}$.

\subsection{Per-UE allocation for system performance guarantees}
\label{sec:scheduling-constraints}
Once the push subframe is over, the \gls{cpu} decodes the received packets obtaining $V_u$ and $B_u$, $\forall u \in\mc{S}$, used to perform \gls{re} assignment. Here, we analyze how to evaluate the minimum amount of \glspl{re} per \gls{ue}, namely $Q_u$, to satisfy both the communication and localization constraints.

\paragraph*{Communication} The aim is obtaining the amount of resources $Q_u^\mathrm{com}$ guaranteeing an achievable rate when \gls{ue} $u\in\mc{S}$ transmits $B_u$ bytes. 
Due to the orthogonal allocation, \gls{ue}'s transmitted symbols are estimated employing \gls{mr} combining to $\mb{Y}_i$, i.e., the \glspl{ap}' combiner vectors is $\mb{v}_{i} = \hat{\mb{h}}_{u,i}$. Using the \emph{\gls{uatf} bound}, an achievable \gls{se} for a single $(f,t)$ in the \gls{re} is~\cite[Th. 5.2]{demir2021cellfree}
\begin{equation} \label{eq:se}
    \rho_u(\bm{\beta}_u) = \log\left(1 +  \frac{p_u}{\sigma_w^2} \frac{M \left|\sum_{\ell\in\mc{L}} \beta_{u,\ell} \right|^2}{ 2\sum_{\ell\in\mc{L}} \beta_{u,\ell} + \frac{\sigma_w^2}{p_u}L }\right). \quad \text{[bit/s/Hz]}
\end{equation}
To deliver a packet of $B_u$ bytes, \gls{ue} $u\in\mc{S}$ thus needs
\begin{equation} \label{eq:res-comm}
    Q_u^\mathrm{com} \ge 8B_u(\rho_u(\bm{\beta}_u)\, (T-1) F)^{-1}.
\end{equation}
We employ an \gls{ed} on the channel estimate obtained in the push subframe to estimate $\bm{\beta}_{u}$. Denoting $i^*\in\mc{P}$ as the \gls{re} in which \gls{ue} $u\in\mc{S}$ transmitted within the push subframe, we have
\begin{equation} \label{eq:beta_hat}
    \hat{\bm{\beta}}_{u} = \frac{1}{F M}\sum_{\ell\in\mc{L}}\lVert \mb{e}_\ell\T\hat{\mb{h}}_{u,i^*}\rVert_2^2\mb{e}_{\ell} \approx \bm{\beta}_{u} + \frac{\sigma_w^2}{p_u T F M} \bm{\chi}_{FM}^2,
\end{equation}
with $\bm{\chi}_{FM}^2\in\mathbb{R}^L$ being a Chi-squared distributed vector with $FM$ degrees of freedom.
Thus, the number of pull subframe's \glspl{re} allowing \gls{ue} $u\in\mc{U}$ to reliably deliver $B_u$ bytes is\footnote{The approximation is tight for high $T F M$, which is usually the case. The \gls{ed} noise may make the rate unachievable, but due to the \gls{uatf} bound and the high $FTM$ value, the chances of this occurrence are negligible.}
\begin{equation}
    Q_u^\mathrm{com} = \left\lceil8B_u(\rho_u(\hat{\bm{\beta}}_u)\, (T-1) F)^{-1}\right\rceil.
\end{equation}

\paragraph*{Localization} The aim is obtaining the amount of resources $Q_u^\mathrm{loc}$ guaranteeing a \gls{peb} $\le \epsilon$ for all \glspl{ue} $u\in\mc{S}$. 
Denoting the set indexing $Q_u^\mathrm{loc}$ as $\mc{Q}_u$, the \gls{speb} is given by $\mathrm{PEB}_u^2 = \tr[\left(\sum_{i\in\mc{Q}_u}\mb{J}_i\right)^{-1}]_{1:2;1:2}$ being $\mb{J}_i$ the \gls{fim} for \gls{re} $i$ over the vector $\bm{\kappa}_u = [\mb{x}_u\T, \bm{\beta}_u\T, \bm{\psi}_u\T, \delta\tau_{u}]\T\in\mathbb{R}^{2L +1}$ collecting the location and nuisance parameters of \gls{ue} $u\in\mc{U}$. It can be bounded by
\begin{equation} \label{eq:peb_ub}
     \mathrm{PEB}_u^2  \le \tr\left[\left(Q_u^\mathrm{loc} \mb{J}_1\right)^{-1}\right]_{1:2;1:2} \le \epsilon^2,
\end{equation}
 by neglecting the delay accuracy gain provided by scheduling over larger bandwidths~\cite{Wymeersch2022fundamentals}, i.e., neglecting spectral diversity and treating all \glspl{re} as time slots. Eq.~\eqref{eq:peb_ub} yields
\begin{equation}
    Q_u^\mathrm{loc} = \left\lceil \epsilon^{-2}\, \tr[\left(\mb{J}_{1}\right)^{-1}]_{1:2;1:2}\right\rceil. 
\end{equation}
To compute the \gls{fim}, we remark that localization method occurs after data decoding, letting use the retrieved symbols $\mb{S}_{u,i}$ as pilots by applying eq.~\eqref{eq:ce} to all the \gls{ofdm} symbols. Thus, we have $T$ independent copies of the estimated channel $\hat{\mb{h}}_{u,i} \sim\mc{CN}(\mb{h}_{u,i}, \frac{\sigma_w^2}{p_u}\mb{I}_{LFM})$, yielding~\cite{kay1993estimation}
\begin{equation}
    \mb{J}_i = 2 T \frac{p_u}{\sigma_w^2} \Re\left\{ \nabla_{\bm{\kappa}_u}  \mb{h}_{u,i}\H \nabla_{\bm{\kappa}_u}  \mb{h}_{u,i} \right\}
\end{equation}
where the gradients are given in the Appendix, which depend on $\mb{x}_u$ and $\bm{\beta}_u$. To approximate them, we use $\hat{\bm{\beta}}_u$ in~\eqref{eq:beta_hat}, and the worst-case position $\tilde{\mb{x}}_u = \arg\min_{\mb{x}} \tr[(\mb{J}_{1}(\mb{x}, \hat{\bm{\beta}}_u)^{-1}]_{1:2;1:2}$, found by grid-search. Due to eq.~\eqref{eq:channel-model-full}, $\tilde{\mb{x}}_u$ depends on $\mb{B}_{\ell}$ only, leading to a single grid-search per scenario.

\paragraph*{ISAC} The minimum amount of \glspl{re} per \gls{ue} satisfying both communication and localization constraints is
\begin{equation} \label{eq:Qu}
    Q_u = \max\{Q_u^\mathrm{com}, Q_u^\mathrm{loc}\}.
\end{equation}

\subsection{Pull subframe scheduling strategy}
\label{sec:mac-pull}
Let us now schedule the pull subframe, remarking that only the \glspl{ue} that managed to transmit during the push subframe, i.e, $u\in\mc{S}$, have communicated their \gls{voi}, and are thus eligible for scheduling.
Thus, we denote as $\mathbf{g}\in\mc{S}^Q$ the scheduling vector, whose element $g_{i} \in \mc{S}$ indicates which \gls{ue} $u\in\mathcal{S}$ is scheduled to pull subframe's \gls{re} $i\in\mc{Q}$. As for Sec.~\ref{sec:scheduling-constraints}, each \gls{ue} $u\in\mc{S}$ has a minimum allocation $Q_u$ given in~\eqref{eq:Qu}, required to satisfy both communication and localization constraints. The utility for the \gls{cpu} from scheduling \gls{ue} $u$ is then equal to $V_u$ if both constraints are satisfied and $0$ otherwise. We then pose the following optimization problem:
\begin{equation}    \label{eq:scheduling-problem} \mb{g}^*=\argmax_{\mb{g}\in\mc{U}^Q}\sum_{u\in\mc{S}} V_u I\left(\sum_{i\in\mc{Q}}I(g_i=u)\geq Q_u\right),
\end{equation}
where $I(a) = 1 \iff a$ is true, 0 otherwise.
This is an instance of the knapsack problem, which belongs to the NP-Hard class, and there are no known polynomial-time optimization algorithms. We then use a heuristic approach: $a)$ we first solve the continuous relaxation of the problem by sorting the \glspl{ue} by their value per \gls{re}, $\frac{V_u}{Q_u}$, and assign the requested \glspl{re} to the first \glspl{ue} that avoids overflow; then, $b$) we adopt a \emph{remove-one local search heuristic}, removing one element from the solution and checking if it leads to an improvement, until the solution is stable. 
While this heuristic have an optimality gap, this should be relatively small in practical scenarios, and it can be run in real time by the \gls{cpu}.


\section{Simulation Settings and Results}
\label{sec:simulations}
\addtolength{\tabcolsep}{-0.5em}
\begin{table}[b]
    \centering
    \vspace{-0.5cm}
    \caption{Simulation parameters.}
    \scriptsize
    \begin{tabular}{ccc|ccc}
    \toprule
    \textbf{Param.} & \textbf{Meaning} & \textbf{Value} & \textbf{Param.} & \textbf{Meaning} & \textbf{Value}\\
    \midrule
    $L$ & no. \glspl{ap} &$4$ & $M$ & no. antenna & $2$ \\
    $U$ & no. \glspl{ue} & $50$ & $v_u$ & \gls{ue} max speed & $20$ km/h \\
    $V_u$ & \gls{voi} & $\mc{U}(0,1)$ & $B_u$ & obs. size & $\mathrm{Geo}(2^{10})$\\
    $N_\mathrm{rb}$ & num. \gls{rb} & $25$ & $N_\mathrm{ts}$ & no. slots & $11$\\
    $F$ & no. subcar. & $12$ & $T$ & \gls{ofdm} sym. & $7$\\
    $\Delta_F$ & subcar. spacing & $30$~kHz & $T_\mathrm{cp}$ & \gls{cp} time & $2.35$~$\mu$s \\
    $N_p$ & push slots & $2$ & $P$ & push \glspl{re} & $50$ \\
    $N_q$ & pull slots & $5$ &
    $Q$ & pull \glspl{re}& $125$\\ 
    $p_u$ & Tx. power & $1$~mW & $\sigma_w^2$ & \gls{awgn} power & $-95$~dBm\\
    $\theta$ & \gls{voi} thres. & $0.7$ & $\epsilon$ & \gls{peb} constraint & $1$~m\\
       \bottomrule
    \end{tabular}    
    \label{tab:param}
\end{table}

We perform numerical simulations considering a square scenario with a side of $200$ m, where the \glspl{ap} and their antennas are symmetrically positioned with antenna spacing $\lambda/2$. Each simulation contains $100$ frames; the \glspl{ue} are randomly positioned and then move following Brownian motion. Large-scale fading $\bm{\beta}_u$ is evaluated through \gls{fspl} with gain $2.15$ dBi. The results are averaged over $10^4$ episodes. The simulation parameters are given in Tab.~\ref{tab:param}.\footnote{The simulation code is available at \href{https://github.com/lostinafro/push-pull-for-isac}{this link}.}

We compare the performance of our \emph{\gls{goia}} with three benchmarks. $1)$ \emph{\gls{poia}}: the first $P$ \gls{ue} with $V_u \ge \theta$, ordered by \gls{voi}, are able to access the push subframe with \emph{no collision}; then, per-\gls{ue} allocation and pull scheduling strategy are performed as \gls{goia}. $2)$ \emph{\Gls{goca}}, which perform the strategies of \gls{goia} without enforcing localization, i.e., $Q_u^\mathrm{loc} = 0$, $\forall u\in\mc{U}$; $3)$ \emph{\gls{viba}}: \glspl{ue} attempt transmission regardless of their \gls{voi}; the \gls{cpu} solves~\eqref{eq:scheduling-problem} by setting $V_u = 1$, $\forall u\in\mc{U}$.

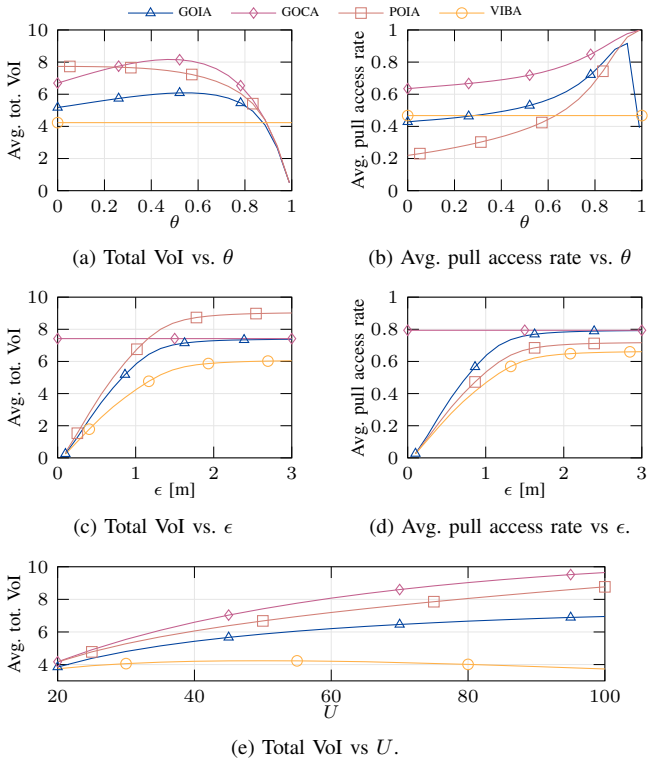
\begin{figure}
\centering \subfloat{\begin{tikzpicture}
\begin{axis}[
    width=0cm,
    height=0cm,
    axis line style={draw=none},
    tick style={draw=none},
    at={(0,0)},
    scale only axis,
    xmin=0,
    xmax=1,
    xtick={},
    ymin=0,
    ymax=1,
    ytick={},
    legend cell align={left},
    legend style={at={(0.5,0.)}, anchor=south, draw=none, fill=none, /tikz/every even column/.append style={column sep=5pt}},
    legend columns = -1,
    ]
\addlegendimage{draw=color0, mark=triangle}
\addlegendentry{GOIA}
\addlegendimage{draw=color1, mark=diamond}
\addlegendentry{GOCA}
\addlegendimage{draw=color2, mark=square}
\addlegendentry{POIA}
\addlegendimage{draw=color3, mark=o}
\addlegendentry{VIBA}

\end{axis}
\end{tikzpicture}}\\
    \vspace{-5.5mm}
    \setcounter{subfigure}{0}
    \subfloat[Total \gls{voi} vs. $\theta$ \label{fig:theta:voi}]{\begin{tikzpicture}
\pgfplotstableread{csv/goia_vs_theta.csv}\goia;
\pgfplotstableread{csv/goca_vs_theta.csv}\goca;
\pgfplotstableread{csv/poia_vs_theta.csv}\poia;
\pgfplotstableread{csv/viba_vs_UE.csv}\viba;

\begin{axis}[
    width=\halfwidth,
    height=\fheight,
    xlabel={$\theta$},
    ylabel={Avg. tot. \gls{voi}},
    xmin=0, xmax=1,
    ymin=0, ymax=10,
    ymajorgrids,
    xmajorgrids,
    xlabel shift=-2mm,
    ylabel shift=-2mm,
]

\addplot[color=color0, mark=triangle, mark repeat=5]
    table[x=Theta,y=avg_voi_tot] {\goia};
\addplot[color=color1, mark=diamond, mark repeat=5, mark phase=1]
    table[x=Theta,y=avg_voi_tot] {\goca};
\addplot[color=color2, mark=square, mark repeat=5, mark phase=2]
    table[x=Theta,y=avg_voi_tot] {\poia};

\addplot[color=color3, mark=o, mark repeat = 2]
table{
0, 4.2324366044721895
0.5, 4.2324366044721895
1, 4.2324366044721895
};

\end{axis}
\end{tikzpicture}}\hfill
    \subfloat[Avg. pull access rate vs. $\theta$ \label{fig:theta:pull}]{\begin{tikzpicture}
\pgfplotstableread{csv/goia_vs_theta.csv}\goia;
\pgfplotstableread{csv/goca_vs_theta.csv}\goca;
\pgfplotstableread{csv/poia_vs_theta.csv}\poia;
\pgfplotstableread{csv/viba_vs_UE.csv}\viba;

\begin{axis}[
    width=\halfwidth,
    height=\fheight,
    xlabel={$\theta$},
    ylabel={Avg. pull access rate},
    xmin=0, xmax=1,
    ymin=0, ymax=1,
    ymajorgrids,
    xmajorgrids,
    xlabel shift=-2mm,
    ylabel shift=-2mm,
]

\addplot[color=color0, mark=triangle, mark repeat=5]
    table[x=Theta,y=avg_pull_access_rate] {\goia};
\addplot[color=color1, mark=diamond, mark repeat=5, mark phase=1]
    table[x=Theta,y=avg_pull_access_rate] {\goca};
\addplot[color=color2, mark=square, mark repeat=5, mark phase=2]
    table[x=Theta,y=avg_pull_access_rate] {\poia};

\addplot[color=color3, mark=o, mark repeat = 1]
table{
0, 0.4671981687056664
1, 0.4671981687056664
};

\end{axis}
\end{tikzpicture}}
    \vspace{-2mm}
    \subfloat[Total \gls{voi} vs. $\epsilon$ \label{fig:epsilon:voi}]{\begin{tikzpicture}
\pgfplotstableread{csv/goia_vs_epsilon.csv}\goia;
\pgfplotstableread{csv/goca_vs_UE.csv}\goca;
\pgfplotstableread{csv/poia_vs_epsilon.csv}\poia;
\pgfplotstableread{csv/viba_vs_epsilon.csv}\viba;

\begin{axis}[
    width=\halfwidth,
    height=\fheight,
    xlabel={$\epsilon$ [m]},
    ylabel={Avg. tot. \gls{voi}},
    xmin=0, xmax=3,
    ymin=0, ymax=10,
    ymajorgrids,
    xmajorgrids,
    xlabel shift=-2mm,
    ylabel shift=-2mm,
]

\addplot[color=color0, mark=triangle, mark repeat=5]
    table[x=epsilon,y=avg_voi_tot] {\goia};
\addplot[color=color1, mark=diamond, mark repeat=1]
    table{
    0, 7.419957802298627
    1.5, 7.419957802298627
    3, 7.419957802298627
    };
\addplot[color=color2, mark=square, mark repeat=5, mark phase=2]
    table[x=epsilon,y=avg_voi_tot] {\poia};

\addplot[color=color3, mark=o, mark repeat = 5, mark phase=3]
table[x=epsilon,y=avg_voi_tot] {\viba};

\end{axis}
\end{tikzpicture}}\hfill
    \subfloat[Avg. pull access rate vs $\epsilon$. \label{figs:epsilon:pull}]{\begin{tikzpicture}
\pgfplotstableread{csv/goia_vs_epsilon.csv}\goia;
\pgfplotstableread{csv/goca_vs_UE.csv}\goca;
\pgfplotstableread{csv/poia_vs_epsilon.csv}\poia;
\pgfplotstableread{csv/viba_vs_epsilon.csv}\viba;

\begin{axis}[
    width=\halfwidth,
    height=\fheight,
    xlabel={$\epsilon$ [m]},
    ylabel={Avg. pull access rate},
    xmin=0, xmax=3,
    ymin=0, ymax=1,
    ymajorgrids,
    xmajorgrids,
    xlabel shift=-2mm,
    ylabel shift=-2mm,
]

\addplot[color=color0, mark=triangle, mark repeat=5]
    table[x=epsilon,y=avg_pull_access_rate] {\goia};
\addplot[color=color1, mark=diamond, mark repeat=1]
    table{
    0, 0.7938851213411755
    1.5, 0.7938851213411755
    3, 0.7938851213411755
    };
\addplot[color=color2, mark=square, mark repeat=5, mark phase=6]
    table[x=epsilon,y=avg_pull_access_rate] {\poia};

\addplot[color=color3, mark=o, mark repeat = 5, mark phase=9]
table[x=epsilon,y=avg_pull_access_rate] {\viba};

\end{axis}
\end{tikzpicture}} \\
    \vspace{-2mm}
    \subfloat[Total \gls{voi} vs $U$. \label{fig:ue}]
    {\begin{tikzpicture}
\pgfplotstableread{csv/goia_vs_UE.csv}\goia;
\pgfplotstableread{csv/goca_vs_UE.csv}\goca;
\pgfplotstableread{csv/poia_vs_UE.csv}\poia;
\pgfplotstableread{csv/viba_vs_UE.csv}\viba;

\begin{axis}[
    width=\fwidth,
    height=0.35\columnwidth,
    xlabel={$U$},
    ylabel={Avg. tot. \gls{voi}},
    xmin=20, xmax=500,
    ymin=0, ymax=12,
    ymajorgrids,
    xmajorgrids,
    xlabel shift=-2mm,
    ylabel shift=-2mm,
]

\addplot[color=color0, mark=triangle, mark repeat=5]
    table[x=U,y=avg_voi_tot] {\goia};
\addplot[color=color1, mark=diamond, mark repeat=5, mark phase=1]
    table[x=U,y=avg_voi_tot] {\goca};
\addplot[color=color2, mark=square, mark repeat=5, mark phase=2]
    table[x=U,y=avg_voi_tot] {\poia};

\addplot[color=color3, mark=o, mark repeat = 5, mark phase=3]
table[x=U,y=avg_voi_tot] {\viba};

\end{axis}
\end{tikzpicture}}\hfill
    \vspace{-1mm}
    \caption{Avg. performance.}
    \vspace{-5mm}
    \label{fig:results}
\end{figure}

Figs.~\ref{fig:theta:voi} and~\ref{fig:theta:pull} show the results vs. the \gls{voi} threshold, in terms of avg. total \gls{voi} of the scheduled \glspl{ue} in the pull subframe, and avg. pull success rate, computed as the ratio between the amount of scheduled \glspl{ue} within the pull subframe and $|\mc{S}|$, respectively. \Gls{poia} performs best for $\theta\le0.25$, even with the lowest avg. pull access rate. Not being affected by collision, only the \glspl{ue} carrying the highest \gls{voi} data are scheduled. \Gls{viba} performs the worst, except when $\theta\approx 1$.
\gls{goia} and \gls{goca} achieve a maximum for $\theta \approx 0.52$ corresponding to a push success rate of $\approx 0.63$. For $\theta>0.52$, a fewer number of \glspl{ue} attempt push access, degrading \gls{voi} performance.  Nevertheless, \gls{goca} outperform \gls{goia}, thanks to a pull access rate of $\approx 0.72$ against $\approx0.53$. The reason is that the burden of resources of localization service is on average higher than the one required for communication due to the relatively high \gls{peb} constraints, and a loose bound in~\eqref{eq:peb_ub}.

Figs.~\ref{fig:epsilon:voi} and~\ref{figs:epsilon:pull} corroborate the previous statement, showing the results vs. the \gls{peb} constraint $\epsilon$. When this value is loose ($2.8$m) \gls{goia} performs as well as \gls{goca}, due to the reduced number of resources needed for localization--on avg. $90\%$ of the time $Q_u^\mathrm{com} \ge Q_u^\mathrm{loc}$@$\epsilon = 2.8$m, vs. $51\%$@ $\epsilon=1$m.


Finally, Fig.~\ref{fig:ue} shows the performance vs. $U$ proving that our framework can improve the total \gls{voi} even when the number of \glspl{ue} are triple than the \glspl{re} for push access $P$, achieving the maximum of $\approx 7.27$ for $U=160$,  while \gls{viba} fails due to high contention. 

\section{Conclusion and Future Work}
\label{sec:conc}
We presented a goal-oriented \gls{isac}-enabled access framework for \gls{dt}. By combining a \gls{voi}-aware push phase with a scheduled pull phase, our approach maximizes the per-frame total \gls{voi}, guaranteeing reliable communication and strict \gls{peb}. Simulation results demonstrate that the proposed \gls{goia} outperforms \gls{voi}-blind schemes, while showing the performance gap due to \gls{isac}. Future works will explore \gls{voi} estimation, tracking, and relax channel modeling assumptions.

\appendix
Implicitly taken w.r.t. $\mb{h}_{u,i}$, the gradients are:
\[
\begin{aligned}
    \nabla_{\mb{x}_u} &= \sum_{\ell\in\mc{L}} \sqrt{\beta_{u,\ell}} e^{\psi_{u,\ell}} \,  (\mb{e}_\ell \otimes \nabla_{\mb{x}_u}\mb{f}_{u,\ell,i} \otimes \mb{a}_{u,\ell}\\
    &\qquad+ \mb{e}_\ell \otimes \mb{f}_{u,\ell,i} \otimes \nabla_{\mb{x}_u}\mb{a}_{u,\ell}) \in\mathbb{C}^{LFM \times 2},
\end{aligned}
\]
\[
\nabla_{\bm{\beta}_u} = \sum_{\ell\in\mc{L}} \frac{e^{j\psi_{u,\ell}}}{2 \sqrt{\beta_{u,\ell}}}(\mb{e}_\ell \otimes \mb{f}_{u,\ell,i} \otimes \mb{a}_{u,\ell}) \mb{e}_\ell\T \in\mathbb{C}^{LFM \times L},
\]
\[
\nabla_{\bm{\psi}_u} = j\sum_{\ell\in\mc{L}} \sqrt{\beta}_{u,\ell}e^{j\psi_{u,\ell}}(\mb{e}_\ell \otimes \mb{f}_{u,\ell,i} \otimes \mb{a}_{u,\ell}) \mb{e}_\ell\T \in\mathbb{C}^{LFM \times L}, 
\]
\[
    \nabla_{\delta\tau_u} = \sum_{\ell\in\mc{L}} \sqrt{\beta_{u,\ell}} e^{j\psi_{u,\ell}} \,  \mb{e}_\ell \otimes \nabla_{\delta\tau_u}\mb{f}_{u,\ell,i} \otimes \mb{a}_{u,\ell} \in\mathbb{C}^{LFM},
\]
where $\nabla_{\mb{x}_u}\mb{f}_{u,\ell, i} = - j \frac{2\pi \Delta_F}{c}  (\mb{n}_{f} \odot \mb{f}_{u,\ell, i})  \frac{(\mb{x}_u - \mb{b}_\ell)\T}{\ltwonorm{\mb{x}_u - \mb{b}_\ell}}$,$  \nabla_{\mb{x}_u}\mb{a}_{u,\ell}\T = j \frac{2\pi}{\lambda } \mb{a}_{u,\ell}\mb{1}_2 \odot  \frac{(\mb{B}_{\ell} -\mb{b}_\ell)\T\mb{A}_\ell(\mb{x}_u)}{\ltwonorm{\mb{x}_u - \mb{b}_\ell}}$,
$\nabla_{\delta\tau_u}\mb{f}_{u,\ell,i} = - j2\pi \Delta_F (\mb{n}_f \odot \mb{f}_{u,\ell,i})$, 
$\mb{n}_f = [f]_{f\in\mc{F}_i}\T$, and $
\mb{A}_\ell(\mb{x}_u) =\mb{I}_2 - \frac{(\mb{x}_u - \mb{b}_\ell)(\mb{x}_u - \mb{b}_\ell)\T}{\ltwonorm{\mb{x}_u - \mb{b}_\ell}^2}$.

\bibliographystyle{bib/IEEEtranNoURL}
\bibliography{bib/IEEEabrv, bib/refs}
\end{document}